\documentclass[prl,twocolumn]{revtex4}
\usepackage{hyperref}
\usepackage[dvips]{graphicx}
\usepackage{latexsym}
\usepackage{verbatim}
\def \be{\begin{equation}}
\def \ende{\end{equation}}
\def \rodd{R_\mathrm{odd}}

\usepackage{xspace,amssymb,amsfonts,amsmath}
\usepackage{color}

\begin{document}

\title{Evidence of the role of contacts on the observed electron-hole asymmetry in graphene}%
\author{B. Huard}
\author{N. Stander}
\author{J.A. Sulpizio}
\author{D. Goldhaber-Gordon}
\affiliation{Stanford University, Department of Physics, Stanford, California, USA}%

\begin{abstract}
We perform electrical transport measurements in graphene with
several sample geometries. In particular, we design ``invasive''
probes crossing the whole graphene sheet as well as ``external''
probes connected through graphene side arms. The four-probe
conductance measured between external probes varies linearly with
charge density and is symmetric between electron and hole types of
carriers. In contrast measurements with invasive probes give a
strong electron-hole asymmetry and a sub-linear conductance as a
function of density. By comparing various geometries and types of
contact metal, we show that these two observations are due to
transport properties of the metal/graphene interface. The asymmetry
originates from the pinning of the charge density below the metal,
which thereby forms a \emph{p-n} or \emph{p-p} junction depending on
the polarity of the carriers in the bulk graphene sheet. Our results
also explain part of the sub-linearity observed in conductance as a
function of density in a large number of experiments on graphene,
which has generally been attributed to short-range scattering only.

\end{abstract}
\maketitle

Graphene, a crystalline monolayer of carbon, has a remarkable band
structure in which low energy charge carriers behave similarly to
relativistic fermions, making graphene a promising material for both
fundamental physics and potential applications
\cite{neto_electronic_2007}. Most interesting predicted transport
properties require that charge carriers propagate with minimal
scattering. Recently experimentalists have succeeded in reducing
disorder \cite{bolotin_2008,andrei_2008} and have shown the
important role of nearby impurities on the mobility of charge
carriers \cite{novikov_elastic_2007,adam_self-consistent_2007}. In
contrast, the effect of metallic contacts on transport has received
little attention in experiments. For instance, most experiments show
a clear difference between the conductances at exactly opposite
densities, a phenomenon previously attributed to different
scattering cross-sections off charged impurities for opposite
carrier polarities\cite{novikov_numbers_2007,chen_charged_2007}. In
this letter, we show that transport properties of the interface
between graphene and metal contacts can also lead to such an
asymmetry. This effect is due to charge transfer from the metal to
graphene leading to a \emph{p-p} or \emph{p-n} junction in graphene
depending on the polarity of carriers in the bulk of the sheet. We
also show that this leads to sub-linear conductance as a function of
gate voltage, which is traditionally attributed to short-range
scattering
\cite{Nomura_2006,Ando_2006,fasolino_intrinsic_2007,katsnelson_electron_2008}.
With a proper measurement geometry, we find conductivity linear in
density up to at least $n=7~10^{12}~\mathrm{cm}^{-2}$ showing that
short range scattering plays a negligible role.

\begin{figure}[hptb]
\begin{center}
\includegraphics[width=8cm]{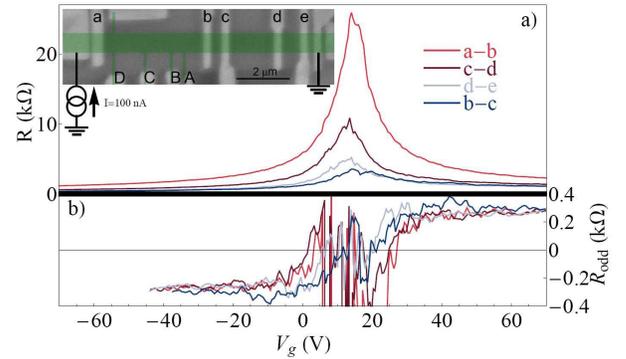}
\caption{\textbf{a)} 4-probe resistance calculated from voltages
measured between invasive probes as a function of gate voltage while
a steady oscillating current of 100~nA runs along the whole graphene
sheet. Inset: Scanning Electron Microscope image of the graphene
sample TiAu1 connected to Hall probes (A-D) and invasive probes
(a-e). For clarity, graphene has been colorized according according
to topography measured by atomic force microscopy. \textbf{b)} Given
the charge neutrality gate voltage $V_g^0$ identified from Quantum
Hall Effect measurements (see text), we plot here the asymmetry
between electrons and holes by showing the odd part of the
resistance defined in Eq.~\ref{eq1}.\label{Fig1}}
\end{center}
\end{figure}

In order to investigate the properties of the graphene/metal
interface, we used two types of metallic voltage probes (see inset
Fig.~\ref{Fig1}). ``Invasive'' probes (like a-e in Fig.~\ref{Fig1})
extending across the full graphene strip width are sensitive to
contact and sheet properties while ``external'' probes (like A-D in
Fig.~\ref{Fig1}) connected to narrow graphene arms on the side of
the strip are sensitive to sheet properties only. All the graphene
samples described in this letter are prepared by successive
mechanical exfoliation of Highly Oriented Pyrolytic Graphite (HOPG)
grade ZYA from General Electric (distributed by SPI) using an
adhesive tape (3M Scotch Multitask tape with gloss finish). The
substrate is a highly n-doped Si wafer, used as a gate (capacitance
13.6~nF.cm$^{-2}$ from Hall effect measurements), on which a layer
of SiO$_2$ 297~nm thick is grown by dry oxidation. Metallic probes
are patterned using standard electron beam lithography followed by
electron beam evaporation of metal (see Table~\ref{Tab1}). Finally,
the graphene sheets are etched in dry oxygen plasma (1:9 O$_2$:Ar)
into the desired shape. The voltage measurements between probes are
performed in liquid Helium at 4~K using a lock-in amplifier at a
frequency between 10 and 150~Hz with a bias current of 100~nA. All
samples were also measured in perpendicular magnetic fields up to
8T, and show the quantum Hall plateaus characteristic of monolayer
graphene. Most samples were additionally characterized by Raman
scattering, in each case showing the typical graphene spectrum(see
Supplementary material \cite{EPAPS}).

Fig.~\ref{Fig1} shows the 4-probe resistances measured between four
pairs of invasive probes in the sample TiAu1, as a function of gate
voltage $V_g$. The resistance is maximal close to the value $V_g^0$
of the gate voltage where the average charge density is zero. In
order to quantify the asymmetry between electron and hole transport,
we plot in Fig.~\ref{Fig1}b the odd part of the resistance defined
as \be \rodd(\Delta V_g)=\frac{1}{2}[R(V_g^0+\Delta
V_g)-R(V_g^0-\Delta V_g)].\label{eq1}\ende We determined the voltage
$V_g^0$ with good precision using the sharp features of resistance
as a function of density in the Quantum Hall regime at 8~T.
\begin{figure}[hbtp]
\begin{center}
\begin{minipage}{.49\linewidth}
\includegraphics[height=4.5cm]{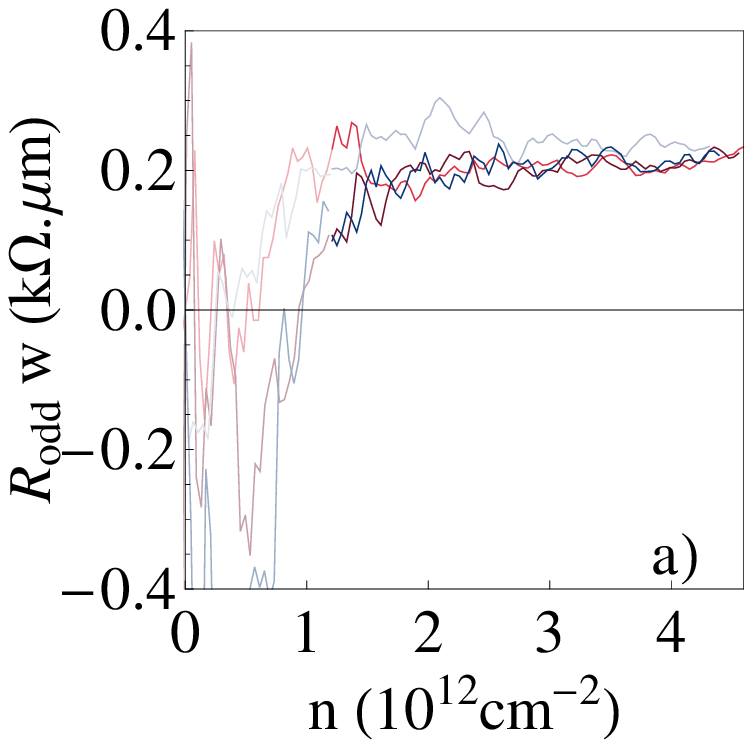}
\end{minipage} \hfill
\begin{minipage}{.49\linewidth}
\includegraphics[height=4.5cm]{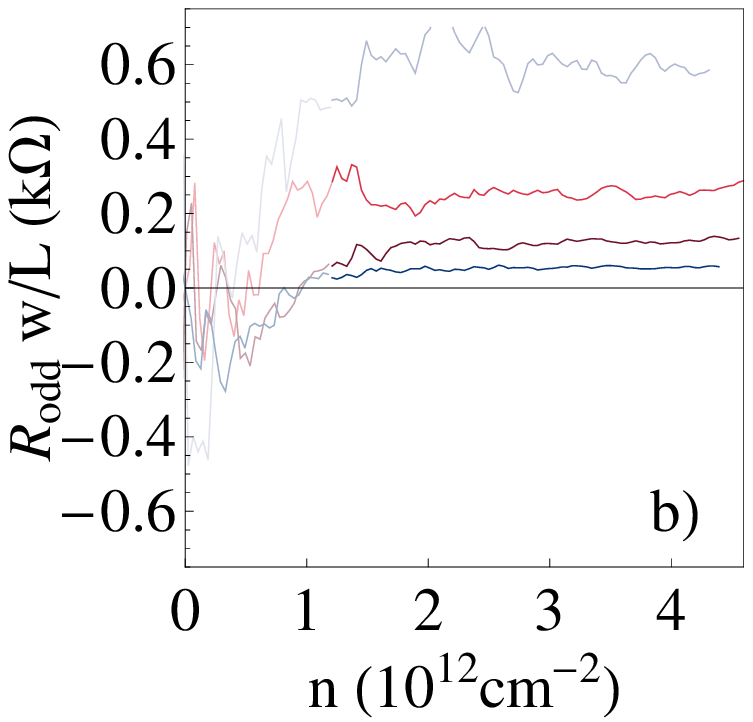}
\end{minipage}
\caption{\textbf{a)} Odd part of resistance normalized by the extent
$w$ of the metal/graphene interface for four pairs of invasive
probes shown in the inset of Fig.~\ref{Fig1} (same colors). The
fluctuating region at densities smaller than
$1.2~10^{12}\mathrm{cm}^{-2}$ has been grayed. The charge density
$n$ is measured using the classical Hall voltage between external
probes, implying a capacitance of $13.6~\mathrm{nF.cm}^{-2}$,
consistent with the measured oxide thickness. \textbf{b)} Same odd
part of the resistance scaled by the ratio of the length $w$ on the
distance $L$ between contacts. \label{Fig2}}
\end{center}
\end{figure}Two regimes of density can be
distinguished. For low densities $n\lesssim
1.2~10^{12}~\mathrm{cm}^{-2}$, $\rodd$ fluctuates widely. The extent
of this fluctuating regime is consistent with the density of charged
impurities $n_i=e(h c_2\mu)^{-1}\approx
0.5~10^{12}~\mathrm{cm}^{-2}$ one would calculate from the
assumption that the mobility $\mu\approx
4600~\mathrm{cm}^2V^{-1}s^{-1}$ (see Fig.~\ref{Fig3}) is dominated
by scattering off charged impurities, where $c_2\approx 0.1$ for
graphene on SiO$_2$
\cite{novikov_elastic_2007,adam_self-consistent_2007}. For larger
densities $n\gtrsim 1.2~10^{12}~\mathrm{cm}^{-2}$, $\rodd$ saturates
to a finite value, corresponding to a higher resistance for
electrons $(V_g>V_g^0)$ than for holes $(V_g<V_g^0)$. Such an
asymmetry was previously predicted and observed in the presence of
charge impurity scattering in graphene
\cite{novikov_numbers_2007,novikov_elastic_2007,chen_charged_2007,adam_self-consistent_2007}.
In that case, the asymmetry comes from a difference between the
scattering cross-sections of positive and negative charge carriers
on a charged impurity. Let us define two different resistivity
functions $\rho_e(|n|)$ for electrons and $\rho_h(|n|)$ for holes as
a function of density $n$. If this is the source of the asymmetry in
resistance, the odd part $\rodd$ should be given by
$2\rodd(n)=(\rho_e(|n|)-\rho_h(|n|))L/w$ for electrons $(n>0)$,
where $L$ is the distance between voltage probes and $w$ is the
width of the graphene strip. However, as can be seen in
Fig.~\ref{Fig2}b, the asymmetry of resistivity inferred in this way
from our four measurements from Fig.~\ref{Fig1} varies widely with
changing $L$. On the contrary, if we associate $\rodd$ with a
specific interface resistance $r(n)$, all curves for different
geometries collapse together (Fig.~\ref{Fig2}a). Therefore, we
propose a more general expression for $\rodd$: \be
\begin{array}{rccc}&\textrm{sheet property}&\ll&\mathrm{interface}\\
2 \rodd(n)= &
\overbrace{[\rho_e(|n|)-\rho_h(|n|)]\frac{L}{w}}&+&\overbrace{r(n)\frac{1}{w}}.\end{array}\label{Eq1}\ende
Repeating the resistance measurements using external probes instead
of invasive probes, we can get rid of the interface term $r(n)/w$ in
Eq.~(\ref{Eq1}) and measure the sheet asymmetry only. To the
precision of our measurements, $\rho_e/\rho_h=1\pm 0.03$ when
averaged on all densities (Fig.~\ref{Fig3}
inset).\begin{figure}[hptb]
\begin{center}
\includegraphics[width=7cm]{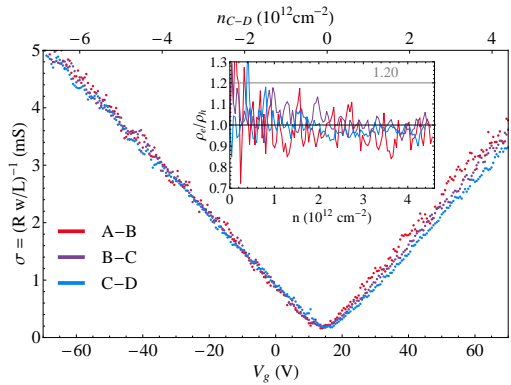}
\caption{Conductivity as a function of gate voltage. Each color
corresponds to a pair of probes identified by two letters on
Fig.~\ref{Fig1}. The slope of these curves corresponds to a mobility
of $\mu\approx 4600~\mathrm{cm}^2V^{-1}s^{-1}$. \textbf{Inset:}
given the charge neutrality gate voltage $V_g^0$ identified from
Quantum Hall Effect measurements (see text), we plot here the ratio
between resistivities for electrons and holes as a function of
carrier density. In contrast to the case of invasive probes, the
average asymmetry is invisible to the precision of our measurement
(note: the observed fluctuations are reproducible). The line
corresponds to the ratio 1.20 observed in
Ref.~\cite{chen_charged_2007} in presence of chemical
dopants.\label{Fig3}}
\end{center}
\end{figure}
The absence of asymmetry between $\rho_e$ and $\rho_h$ is in
contrast with the ratio of about 1.20 Chen \emph{et al.} observed
\cite{chen_charged_2007} when graphene was exposed to chemical
dopants. To understand this apparent discrepancy, let us consider
the three proposed sources of scattering in graphene: short-range
scatterers, charged impurities and corrugation in the graphene
sheet. First, short-range scatterers add a term $\rho_s$ almost
independent of $n$ to the resistivity. From Fig.~\ref{Fig3}, we can
set an upper bound $\rho_s<15~\Omega/\square$ surprisingly small
compared to other reported values \cite{morozov_2008}. Charged
impurities naturally lead to the observed linear dependence of
conductivity on $n$, whereas corrugation requires a particular
height correlation function to give the same behavior, which is thus
less likely \cite{Nomura_2006,
Ando_2006,fasolino_intrinsic_2007,katsnelson_electron_2008}. As has
been predicted and shown experimentally, scattering off charged
impurities of a given polarity occurs at a different rate for
electrons and holes \cite{novikov_numbers_2007,chen_charged_2007},
and it also shifts the voltage $V_g^0$ and decreases the mobility.
However, both in our measurements and in those of
Ref.~\cite{chen_charged_2007} prior to doping, there is no asymmetry
in the resistivity. This could be due to some equilibration between
impurities of opposite polarities, but in this case, the difference
in $V_g^0$ between the experiments is somewhat surprising and would
be worthy of further investigation.

As we have seen, for invasive probes, $\rodd$ scales inversely with
the extent $w$ of the metal/graphene interface. Metallic probes in
contact with graphene are expected to pin the charge density
$n_\mathrm{c}$ in the graphene below the metal thereby creating a
density step along the graphene strip
\cite{golizadeh-mojarad_effect_2007,heinze_carbon_2002,giovannetti_2008}.
The height of this step and the sign of $n_c$ depends on the
mismatch between the work functions of the metal and the graphene
sheet. As we will see, for our choices of contact metal the charge
density in graphene is pinned to a negative value $n_\mathrm{c}$
(\emph{p}-type) below the metal. Thus depending on the polarity of
the carriers in bulk graphene sheet a \emph{p-n} junction or a
\emph{p-p} junction develops close to the metal/graphene interface.
We have shown elsewhere \cite{huard_transport_2007,nimrod} that the
resistances associated with these junctions for opposite values of
the charge density $n$ in the sheet differ by an amount
$r_{n_\mathrm{c}}(n)/w$ where $r_{n_\mathrm{c}}$ depends only on
$n_\mathrm{c}$ and on the length over which the density varies
across the junction
\cite{cheianov_selective_2006,fogler_effect_2008,katsnelson_chiral_2006,zhang_nonlinear_2007}.
This is consistent with the observed positive $R_\mathrm{odd}$; with
\emph{n}-type graphene below the contact one should observe a
negative $R_\mathrm{odd}$. If we further assume that the density
changes from $n_\mathrm{c}$ to $n$ on a very short scale compared to
$(|n_\mathrm{c}|+|n|)^{-1/2}$, we obtain an analytical expression
for $r_{n_\mathrm{c}}$ \cite{EPAPS}. If this is the origin of the
observed asymmetry, $\rodd$ should counter-intuitively decrease when
the mismatch between metal and graphene work functions increases.
The limit where $n_\mathrm{c}$ goes to infinity gives the lowest
possible value of $\rodd$ in this sharp-junction approximation
\cite{EPAPS} \be r_{n_\mathrm{c}}(n)>
2.06\frac{h}{4e^2}n^{-1/2}.\label{Eq2}\ende
\begin{figure}[hptb]
\begin{center}
\includegraphics[width=7cm]{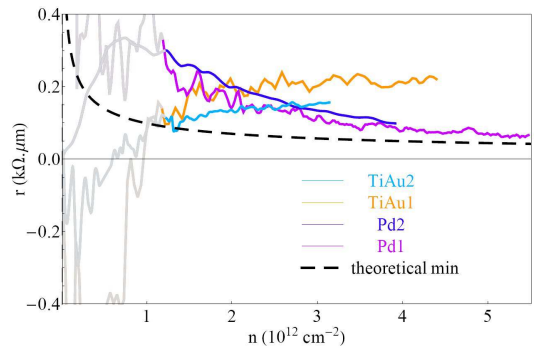}
\caption{Odd part $\rodd$ of the resistance scaled by the inverse
width $w^{-1}$ for various samples and metals described in
Table~\ref{Tab1}.\label{Fig4}}
\end{center}
\end{figure}

\begin{table}[hbtp]
\begin{center}
\begin{tabular}{|c|c|c|}\hline&&\\[-0.25cm]
Sample & Metal thickness & $w~(\mu\mathrm{m})$ \\\hline &&\\[-0.25cm]
Pd1 & Pd(30 nm) & 0.4   \\
Pd2 & Pd(30 nm) & 0.9   \\
TiAu1 & Ti(5 nm)/Au(25 nm) & 0.8   \\
TiAu2 & Ti(3 nm)/Au(15 nm) & 2.4   \\
\hline
\end{tabular}
\caption{Geometrical properties of the samples corresponding to
Fig.~\ref{Fig4}. The measurements shown on
Figs.~\ref{Fig1},\ref{Fig2},\ref{Fig3} were performed on TiAu1. The
type of metal used as a probe and its thickness is given here
together with the length $w$ of the graphene/metal interface.
\label{Tab1}}
\end{center}
\end{table}
On Fig.~\ref{Fig4}, we show the function $r_{n_\mathrm{c}}$ measured
in several graphene sheets contacted with two types of metal (see
Table~\ref{Tab1}). For Pd, which is expected to have a high work
function ($\Phi_\mathrm{Pd}\approx
5.1~\mathrm{eV}<\Phi_\mathrm{graphene}=4.5~\mathrm{eV}$ with the
prediction for graphene, see e.g. Ref.~\cite{giovannetti_2008}), the
function $r_{n_\mathrm{c}}$ is very close to the lower bound
Eq.~(\ref{Eq2}). In contrast, for Ti covered with a layer of Au,
where the work function mismatch should be smaller
($\Phi_\mathrm{Ti}\approx 4.3~\mathrm{eV}$ and
$\Phi_\mathrm{Au}\approx 5.1~\mathrm{eV}$), the function
$r_{n_\mathrm{c}}$ was larger at high densities $n$, suggesting that
the densities $n_\mathrm{c}$ and $n$ are of the same order of
magnitude. We notice that for Pd, $r_{n_\mathrm{c}}$ decreases with
$n$ whereas for Ti/Au it increases, but it is hard to explain this
increase since it would require knowing the potential profile close
to the lead. Finally, as expected\cite{giovannetti_2008}
($\Phi_\mathrm{Al}\approx 4.2~\mathrm{eV}<\Phi_\mathrm{graphene}$),
Ti/Al probes lead to the opposite doping: $R_\mathrm{odd}$ that we
estimate from other works \cite{Heersche_2007,philippe} is negative.

\begin{figure}[hptb]
\begin{center}
\includegraphics[width=8.4cm]{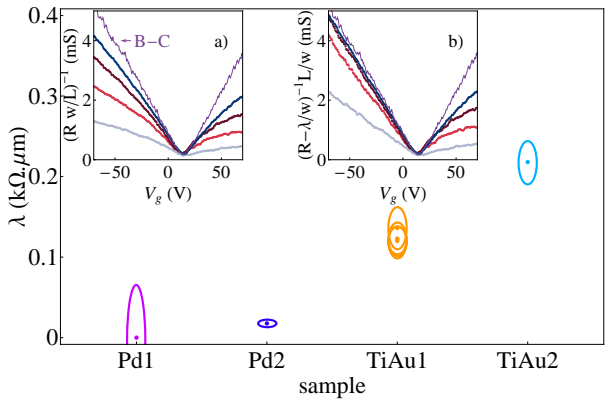}
\caption{\textbf{a)} From the resistance curves plotted in
Fig.~\ref{Fig1}, we show the conductance scaled by the ratio $w/L$.
The non-invasive measurement between probes B and C from
Fig.~\ref{Fig3} is plotted as a thin line for reference. \textbf{b)}
subtracting $\lambda=0.135~\mathrm{k}\Omega.\mu\mathrm{m}$ divided
by the length $w$ of the metal/graphene interface, each curve from
a) is linearized for the \emph{p-}type carriers $(V_g<V_g^0)$.
\textbf{Main panel:} for each four-probe measurement on the samples
from Table~\ref{Tab1}, we plot here the specific resistance
$\lambda$ which best linearizes the conductance as a function of
gate voltage (see text). The best fit is obtained at the dot and the
vertical size of the corresponding ellipse represents the
uncertainty on $\lambda$. \label{Fig5}}
\end{center}
\end{figure}

Charge transfer from the metallic probes has yet another observable
effect on transport. On Fig.~\ref{Fig5}a, we show the conductance
measured using invasive probes scaled by the geometrical aspect
ratio of each section. Even on the hole side $(V_g<V_g^0)$ where
there is no \emph{p-n} junction, a sub-linearity is striking when
compared to the external probe measurement shown in the same figure.
We find that there is a constant specific contact resistance
$\lambda$ such that $(R-\lambda/w)^{-1}$ is linear in density in the
hole region (see Fig.~\ref{Fig5}b). This contact resistance
independent of density $n$ can be attributed to a higher
concentration of short-range scatterers near the contact (perhaps
due to e-beam exposure during lithography) and/or to the region of
constant density $n_\mathrm{c}$ near the contacts. In order to
determine which effect is dominant, we compare the value of
$\lambda$ for the two different metals of Table~\ref{Tab1}.
We find that this resistance is small for Pd probes compared to
Ti/Au probes, consistent with sub-linearity coming from a region of
larger constant density $n_\mathrm{c}$ for Pd than for Ti/Au, and
not from short-range scatterers.

In conclusion, we have shown that all measurements using invasive
metallic probes should exhibit an asymmetry between hole and
electron conductances due to charge transfer at the graphene/metal
interface. Similarly, invasive probes lead to a sub-linearity in the
conductance as a function of density, even in a 4-probe geometry. In
every experiment using invasive probes, one should consider these
effects in the calculation of the conductivity from the resistance
measurement and sample geometry. External probes do not have this
issue and reveal a conductance linear in density.

We thank D. Novikov, M. Fogler and J. Cayssol for enlightening
discussions. This work was supported by the MARCO/FENA program and
the Office of Naval Research \# N00014-02-1-0986. N.~Stander was
supported by a William R. and Sara Hart Kimball Stanford Graduate
Fellowship, J.A.~Sulpizio by a National Science Foundation graduate
fellowship. Work was performed in part at the Stanford
Nanofabrication Facility of NNIN supported by the National Science
Foundation under Grant ECS-9731293. Critical equipments were
obtained on Air Force Grant.


\end{document}